%% file: chang_sol.tex
\newif\ifLNCS
\LNCSfalse
\newif\ifnotLNCS
\notLNCStrue

\ifLNCS
\RequirePackage{amsmath}
\fi
\documentclass{article}
\usepackage{graphicx}

\ifLNCS
\usepackage{makeidx}  
\fi

\usepackage{times}
\usepackage{xcolor}
\usepackage{soul}
\usepackage[utf8]{inputenc}
\usepackage[small]{caption}

\input{macros}

\title{Attaining Equilibria Using Control Sets}

\author{Gleb Polevoy \and
Jonas Schweichhart}

\begin{document}

\ifLNCS
%
%
\fi

\ifLNCS
\fi
\maketitle 



\input{abstract}

\input{introduction}
\input{model}
\input{rel_work}
\input{gen_game}

\input{coord_game_graph}
\input{conclusion}




\bibliographystyle{unsrt}
\bibliography{library}


\end{document}

%% file: macros.tex
\usepackage[ruled]{algorithm2e}

\SetAlFnt{\small}
\SetAlCapFnt{\small}
\SetAlCapNameFnt{\small}
\SetAlCapHSkip{0pt}

\usepackage{url}

\usepackage{latexsym,amsmath,amssymb}
\ifnotLNCS
\usepackage{amsthm}
\fi
\usepackage{graphicx}

\usepackage{ifthen}
\usepackage{mdwlist,paralist}

\usepackage{comment}
\excludecomment{longversion}

\usepackage{epsfig}
\usepackage{epsf}

\usepackage[geometry]{ifsym}
\usepackage{rotating}

\usepackage{float}

\usepackage{tikz}
\usetikzlibrary{positioning,calc}

\newboolean{Longversion} \setboolean{Longversion}{false}

\newcommand*{\sectn}[1]{Section~#1}
\newcommand*{\sectnref}[1]{\sectn{\ref{#1}}} 

\newcommand{\anote}[1]{}

\newcommand{\argmin}{\ensuremath{\mbox{argmin}}} 
 
\newcommand{\set}[1]{\left\{ #1 \right\}} 
 
\newcommand{\paren}[1]{\left( #1 \right)}

\newcommand{\abs}[1]{\left| #1 \right|}

\newcommand{\naturals}{\mathbb{N}} 

\newcommand{\reals}{\mathbb{R}}
\newcommand*{\realsP}{{\ensuremath{\mathbb{R}_{+}}}}

\newcommand*{\harder}[1]{#1-hard}

\newcommand*{\Pol}{\textrm{P}}
\newcommand*{\NP}{\textrm{NP}}
\newcommand*{\coNP}{\textrm{Co-NP}}

\newcommand*{\NPH}{\mbox{\textrm{\harder{\NP}}}}

\newcommand{\calC}{\mathcal{C}}

\newcommand{\calP}{\mathcal{P}}

\newcommand{\calS}{\mathcal{S}}

\newcommand*{\defas}{\ensuremath{\stackrel{\rm \Delta}{=}}}

\DeclareMathOperator*{\Neighb}{N}

\ifnotLNCS
\newtheorem{theorem}{Theorem}
\fi
\newtheorem{defin}{Definition}
\ifnotLNCS
\newtheorem{remark}{Remark}
\newtheorem{lemma}{Lemma}
\newtheorem{proposition}{Proposition}

\newtheorem{corollary}{Corollary}
\fi
\newtheorem{observation}{Observation}
\ifnotLNCS
\newtheorem{example}{Example}
\newtheorem{assumption}{Assumption}

\fi

\newcommand{\defined}[1]{{\bf\emph{#1}}}

\newcommand{\NE}{\text{NE}\xspace} 

\DeclareMathOperator{\con}{conNum} 
\DeclareMathOperator{\dcon}{dirConNum} 
\newcommand{\dConSet}{DCSet} 
\newcommand{\OrdIndDConSet}{OrdInd\dConSet} 
\newcommand{\verOrdIndDConSet}{ver\OrdIndDConSet} 
\DeclareMathOperator{\disc}{disc} 



%% file: abstract.tex
\begin{abstract}
Many interactions result in a socially suboptimal equilibrium,
or in a non-equilibrium state, from which arriving at an equilibrium through simple dynamics can be impossible of too long.
Aiming to achieve a certain equilibrium, we persuade, bribe, or coerce a group of 
participants to make them act in a way that will motivate the rest of the players to act
accordingly to the desired equilibrium.
Formally, 
we ask which subset of the players can adopt the goal equilibrium
strategies that will make acting according to the desired equilibrium
a best response for the other players. We call such a subset a
direct control set, prove some connections to strength of equilibrium, and study the hardness to find such lightest sets, even approximately. We then solve important subcases and provide approximation algorithms, assuming monotonicity.
Next, we concentrate on potential games and
prove that, while the problem of finding such a set is \NP-hard,
even for constant-factor approximation,
we can still solve the problem approximately or even precisely in relevant
special cases. We approximately solve this problem for singleton
potential games and treat more closely specific potential games,
such as symmetric games and
coordination games on graphs.
\end{abstract}

%% file: introduction.tex
\section{Introduction}

The way to predict what will happen in an interaction is studying the set of solutions of the modelling game, such as Nash equilibria~\cite{Nash51}.
Consider a society that has stabilised in an equilibrium that we dislike, such as
high corruption and violence levels or little cooperation, while we can describe a stable
state we would like to have instead of the current one. 
Alternatively, consider a situation being temporarily unstable, and we are not predisposed to wait till it reaches stability through some natural dynamics, or we are not in favour of the equilibrium that can be thus reached. The arising question is then how to bring the society to the preferred stable state. 

There exist several approaches in the spirit of mechanism design
to influence the equilibria of a game, such as announcements that make
a game without equilibria have 
equilibria~\cite{GrantKrausWooldridgeZuckerman2014}, or taxation schemes~\cite{EndrissKrausLangWooldridge2011}. When the planner knows the players' utilities, she can influence the game by offering payments~\cite{AnshelevichSekar2014}, or subsidies and taxes~\cite{PolevoyDziubinski2022}.
All the above methods intervened with the structure of the game, which can be hard to implement. Aiming to arrive at a simpler method,
namely making the agents switch to the desired
equilibrium by controlling the behaviour of a subset of the agents, we
model such an influence by
assuming we can
control some people so that they do what we ask them to, and we would like their actions
to influence the whole society. The controlled players themselves do not have to find the equilibrium behaviour at beneficial at first; after the rest comply and play the desired equilibrium, the equilibrium behavior becomes optimum for everyone, but no necessarily better than the original profile, so there is no optimisation like Stackelberg leader(s) do. The controlled players simply do what they are told to, and after the others best respond, the controlled behavior becomes optimum.
Knowing how many people we have to control and whom exactly
we have to control to practically bring upon the desired change
is important for improving any interaction with multiple
stable states, be it relationships in families, companies, or countries.

After presenting the model in \sectnref{Sec:model}, 
and studying basic properties in \sectnref{Sec:basic_prop},
we consider another reasonable model, which we prove to be intractable even to verification in \sectnref{Sec:ord_ind}, so we keep working on the model of direct control set.
We position our work in the literature in \sectnref{Sex:rel_work} and provide a connection between an equilibrium having a direct control set and its strength in \sectnref{Sec:cntrl_vs_ne}.\footnote{We define the
model for a general solution concept, though we concentrate on Nash
equilibria~\cite{Nash51}.}
We then prove that problem is extremely hard, even for approximation, in \sectnref{Sec:hard}, proceeding to solve a particular case in \sectnref{Sec:DP_tree}, and providing approximation algorithms for the case the problem exhibits monotonicity in \sectnref{Sec:mono_alg}.
We proceed to analyse the important case of potential games in \sectnref{Sec:pot}, especially concentrating on singleton congestion games in \sectnref{Sec:pot:single_cong},
and we then delve into important classes of potential games,
which prove to be amenable to analysis. In particular,
we optimally solve symmetric games with decreasing costs in~\sectnref{Sec:pot:symm}. Next,
we consider coordination games on
graph~\cite{AptdeKeijzerRahnSchaferSimon2017} in \sectnref{Sec:coord_game_graph}, where the idea of
influencing the way agents play though some neighbours becomes especially evident.

We thus model a way to switch from one Nash equilibrium to another
one internally, i.e.~by controlling some agents, and analyse how this can be done efficiently.

%% file: model.tex
\section{Model}\label{Sec:model}

Consider a strategic form \defined{game}
$G = (N, S = S_1 \times S_2 \times \ldots \times S_n, (u_i)_{i = 1, \ldots, n})$,
where $N = \set{1, \ldots, n}$ is the set of
agents, $S_i$ is agent $i$'s strategy set and
$u_i \colon S \to \reals$ is agent~$i$'s utility function.
The \defined{solutions}, forming a \defined{solution set}, are a set of strategy profiles $D \subseteq S$,
defined for a game by a rule called \defined{a solution concept}.
Throughout the whole paper we assume the solution concept to be
\defined{Nash equilibria}, denoted $\NE$,\footnote{In text, \NE{} also abbreviates the words
``Nash  equilibrium(a)'', when no confusion with the notation of the set arises.} unless explicitly said otherwise.
A Nash equilibrium~\cite{Nash51} of $G$ is a strategy
profile $s = (s_1, \ldots, s_n) \in S_1 \times \ldots \times S_n = S$
such that
\begin{equation}
\forall i \in N, \forall s_i' \in S_i: u_i(s) \geq u_i(s_i', s_{-i}),
\end{equation}
where $s_{-i} \defas (s_1, \ldots, s_{i - 1}, s_{i + 1}, s_n)$.

We now formalise the idea of incentivising playing a solution by controlling a
subset of the agents.
\begin{defin}\label{def:dir_bring_sol_to sol}
Consider a profile $s \in S$, a solution concept $D \subseteq S$ and a profile
$d \in D$; for example, $d$ can be a Nash equilibrium.
A set of agents $A \subseteq N$ \defined{brings $s$ to $d$ (directly)} if 
these agents can play $d$ while all others play $s$, so that
for each agent outside $A$, playing $d$ would be a best response (perhaps one from many). Formally, given the profile
$(d_i)_{i \in A}, (s_j)_{j \in N \setminus A}$, for each agent
$i \in N \setminus A$, playing $d_i$ is a best response. \\
This set $A$ is called a \defined{direct control set} with respect to $G$,
$s$ and $d$. \\
\end{defin}
We call the control set direct, because the strategies
that the agents in $A$ use are their strategies in the destination profile $d$, while one could also imagine controlling the other players by playing other strategies,
so that to complete the movement to solution $d$, the agents in $A$ need to switch to $d$ themselves, once the rest best respond by playing
their strategies in $d$. Such a general control set may be sometimes very potent, but this makes the analysis even more complicated and the assumptions about what we can control become stronger, so we assume the direct control by default.

We aim to find the agent sets that can bring from
one solution to another one, especially smallest such sets, or, if the
agents are weighted, such sets of the smallest total weight, in polynomial time. Formally, we define the following problem.
\begin{defin}\label{def:dir_control_set_prob}
The \defined{direct control set} problem receives as input a game
$G = (N, S, (u_i)_{i = 1, \ldots, n})$,
a profile $s$ and a solution profile $d$ of this game, and a weight
function on the agents, $w \colon N \to \realsP$. \\
A \defined{solution} of the direct control set problem is a set of agents $A \subseteq N$.
A \defined{feasible solution} is a solution that brings $s$ to $d$ (directly),
i.e.~a direct control set with respect to $G, s$ and $d$. \\
We aim to find a feasible solution $A$, the objective being its total weight
$w(A) \defas \sum_{a \in A}{w(a)}$, which we aim to minimise.
\end{defin}

The \defined{decision version} of such a problem is an instance with a
natural number $k$, and the question is whether there exists a direct control set of size or weight at most $k$.

\begin{remark}[Manageable Representation]
A naive approach can be providing $n$ utility $n$-dimensional matrices,
storing all the utilities, but that would take $\Omega(2^n)$ space, rendering the brute-force algorithm, testing all the possible $2^n$ subsets of players for being direct control sets, polynomial. This is impracticable, however, to manage such a huge input, so we assume a utility oracle instead.

Formally, we assume the game is represented in $\Theta(\sum_{i \in N}{\abs{S_i}})$ space, explicitly storing the strategies of each player,
while the utilities are provided by a polynomial oracle, which receives a strategy profile~$s$ and a player~$i$ as input, and returns $u_i(s)$ in polynomial time. This resembles the approach of Papadimitrious and Roughgarden~\cite{PapadimitriouRoughgarden2008}.
In some cases, such as for congestion games or polymatrix games, such an algorithm is implied by the definition of the game.

Therefore, our algorithms need to work with such an input, and when reducing another problem to the direct control set, we are required to present a polynomial algorithm for utility calculation, implementing this oracle. 
\end{remark}

\subsection{Basic Properties}\label{Sec:basic_prop}

We start by studying basic properties of control sets
in game with weakly dominant strategies and in two-player
constant-sum games, and then provide general bounds of 
control numbers.
%
%
%
%
First, we consider simple cases.
\begin{remark}
When the desired solution $d$ is weakly dominant,
then the empty set is a (direct) control set. Thus, then
$\con(G, s, d) = \dcon(G, s, d) = 0$.
\end{remark}

\begin{remark}
Two-player constant-sum game has 
$\con(G, s, t) = \dcon(G, s, d) = 0$, when $s, d \in \NE$. 
Indeed, every such profile consists of maxmin strategies,
and any maxmin is a best response to any maxmin of the other player.

When in the starting profile~$s$, at least one player does not play 
a maxmin strategy, then the singleton containing such a player
is a direct control set.
\end{remark}

We next prove that for any $p$, there always exists a game
with any number of players that exceeds $p$ such that its
control number is exactly $p$.
\begin{proposition}\label{prop:game_exist_for_ctrl}
For any natural numbers $n > p \geq 1$, there exists a game $G$
with $n$ players, possessing
equilibria $s$ and $d$, such that
$\dcon(G, s, d) = p$.
\end{proposition}
\begin{proof}
Consider a symmetric game with $n$ players $N$, each player with the
strategy set $S = \set{a, b}$ and the following utility function for 
every player~$i$:
\begin{equation*}
\begin{cases}
	1 & \text{if } s_i = a \text{ and } \abs{\set{j \in N : s_i = s_j}} \geq p + 1, \\
	1 & \text{if } s_i = b \text{ and } \abs{\set{j \in N : s_i = s_j}} \geq p + 1, \\
	0 & \text{otherwise. } \\
\end{cases}
\end{equation*}

This game possesses the Nash equilibrium where everyone plays
$a$ and the equilibrium when all the players play $b$. Being in the equilibrium where all
play $a$, no $q < p$ players can make moving to play $b$ profitable to the
others, because being the $q + 1$th still leaves
one with $0$ utility.

On the other hand, if any $q \geq p$ agents play $b$, then 
after one more player starts playing $b$, there
will be at least $p + 1$ players playing $b$,
yielding that player the utility of $1$.
\end{proof}

In particular, we derive the following.
\begin{corollary}
For any $n \geq 2$, there exists a game
$G = (N, S, u_1, \ldots, u_n)$ with $\abs{N} = n$ agents with
Nash equilibria $s$ and $d$, such that no set of agents smaller
than $n - 1$ can bring $s$ to $d$.
\end{corollary}

By definition of \NE, we observe the following.
\begin{observation}
For any game $G$ and two equilibria $s$ and $d$,
any set of size $\abs{N} - 1$ can bring $s$ to $d$
directly, i.e.~$\dcon(G, s, t) \leq \abs{N} - 1$.
\end{observation}
\begin{proof}
By the definition of a Nash equilibrium,
since $d$ is an equilibrium, setting any $n - 1$ players to play
their strategies in $d$ would make for the remaining player playing
her strategy in $d$ a best response.

\end{proof}

\subsection{Order-Independence}\label{Sec:ord_ind}

We now propose a much stronger requirement on the control set
and prove that is too strong, so that even verifying the feasibility
of a solution becomes intractable.

Direct control assumes some synchronicity of the reaction
of the players to the strategies of the control set, namely, once the players in the control set play $d$, every other player sees that switching to $d$ is a best response and switches to $d$, regardless the possibly ongoing switches of some fellow players. That is relevant in voting, legal agreements, bidding with incomplete information, and other games with incomplete information or the ability to commit to the switch. In other cases, if we wanted to guarantee the motivation to switch holds even after some of the fellow uncontrolled players have switched, we would need to make the following stronger requirement.

\begin{defin}\label{def:dir_bring_sol_to sol_ord_ind}
Consider a profile $s \in S$, a solution concept $D \subseteq S$ and a profile
$d \in D$; for example, $d$ can be a Nash equilibrium.
A set of agents $A \subseteq N$ \defined{order-independently brings $s$ to $d$ (directly)} if 
these agents can play $d$ while all others play $s$, so that
for each agent outside $A$, playing $d$ would be a best response, regardless which subset of its fellow agents of A have switched to $d$ already. Formally, 
for any subset $Y \subseteq N \setminus A$, 
in the profile
$(d_i)_{i \in A}, (d_i)_{i \in Y}, (s_j)_{j \in N \setminus A}$, for each agent
$i \in N \setminus (A\cup Y)$, playing $d_i$ is a best response. \\
This set $A$ is called an \defined{order-independent direct control set} with respect to $G$,
$s$ and $d$. \\
\end{defin}

The problem to be solved is as follows.
\begin{defin}\label{def:dir_control_set_prob:ord_ind}
The \defined{order-independent direct control set} problem receives as input a game
$G = (N, S, (u_i)_{i = 1, \ldots, n})$,
a profile $s$ and a solution $d$ of this game, and a weight
function on the agents, $w \colon N \to \realsP$. \\
A \defined{solution} of the problem is a set of agents $A \subseteq N$.
A \defined{feasible solution} is a solution that order-independently brings $s$ to $d$ (directly),
i.e.~an order-independent direct control set with respect to $G, s$ and $d$. \\
The objective function of a feasible solution $A$ is its total weight
$w(A) \defas \sum_{a \in A}{w(a)}$, which we aim to minimise.
\end{defin}

This problem includes searching for a set that would motivate the players to move to the destination profile, regardless whatever set has moved to the destination profile already. Thus, even given a solution, verifying its feasibility seems hard. We now substantiate this intuition.
\begin{defin}
Given a game $G = (V, E)$, starting profile $s$, destination profile $d$,
and a direct control set $A \subset V$, we call the problem of deciding whether $A$ is order-independent a \verOrdIndDConSet.
\end{defin}
\begin{theorem}
\verOrdIndDConSet{} is a \coNP-complete problem.
\end{theorem}
\begin{proof}
We equivalently prove that the complement problem, namely $\neg\verOrdIndDConSet$, is \NP-complete.

Belonging to the class \NP{} stems from the possibility to non-deterministically guess a subset $Y$ of $V \setminus A$ and check whether there is a player $i \in N \setminus (A \cup Y)$ such that in the profile
$(d_i)_{i \in A}, (d_i)_{i \in Y}, (s_j)_{j \in N \setminus A}$,
playing $d_i$ is not a best response.
Alternatively, an input belongs to $\neg\verOrdIndDConSet$ if and only is there exists a polynomially sized witness $Y$ of $V \setminus A$, such that
in the profile
$(d_i)_{i \in A}, (d_i)_{i \in Y}, (s_j)_{j \in N \setminus A}$,
playing $d_i$ is not a best response, which is polynomially testable.

We carry out the proof of \NP-hardness of $\neg\verOrdIndDConSet$ by providing the following Karp reduction from the \NP-hard problem Dominating Set~\cite{BarYehudaMoran1984} to $\neg\verOrdIndDConSet$. Given an instance of Dominating Set decision problem $(G = (V, E), k)$, 
where $V = \set{1, 2, \ldots, n}$, we define the following game on players $N \defas V \dot\cup \set{0}$. Each player possesses the binary strategy set $S_i = \set{0, 1}$. Each player besides $0$ has the utility identically equal to zero, namely $u_j \equiv 0, \forall j \in V$. Player $0$ has the following utility function:
\begin{equation}
u_0(x) \defas
 \begin{cases}
    0, &    \textbf{if} \text{ in } G, \set{i \in V : s_i = 1} \text{is a dominating set,}\\ 
    &       \textbf{ and} \abs{\set{i \in V : s_i = 1}} \leq k,\\ 
    &       \textbf{ and } s_0 = 1;\\
    1, &    \textbf{otherwise.} \\
  \end{cases}
\end{equation}
Finally, we want to bring strategy profile $s \defas (0, 0, \ldots, 0)$
to $d \defas (1, 1, \ldots, 1) \in \NE$.

The reduction is polynomial, since determining whether a given set is dominating is polynomial. Strategy profile $s \defas (0, 0, \ldots, 0)$ is an \NE, because playing zero is weakly dominant, and $d \defas (1, 1, \ldots, 1) \in \NE$ because for players besides $0$, any strategy yields the same utility, while for player $0$ strategy $1$ is not worse than strategy $1$, because $\abs{\set{i \in V : s_i = 1}} = n$, and we can assume $n > k$, because otherwise the given instance of dominating set is always true, and we can reduce it to any true instance.

Here, the direct control set is $A \defas \emptyset$.
Finally, set $F \subset V$ is a dominating set of size at most $k$ in $G$ if and only if playing $0$ is strictly suboptimal for player $0$, when all the players in $F$ play $1$, and the other $V \setminus F$ play $0$.
This implies the correctness of the above reduction.
\end{proof}

Since even verification is intractable, we do not consider order-independent direct control set problem in this paper, leaving the task of tackling that problem as is or defining a relevant but simpler problem to the future.
\anote{Is Order-Independent direct control set $NP^{NP}$-hard? It's clearly in $NP^{NP}$, but what about hardness?}

%% file: rel_work.tex
\section{Related Work}\label{Sex:rel_work}

\anote{Jonas, please fill this in. Here is a sketch, which converges to what we do, though you're welcome to improve it too.}

Applications of motivating to move to the desired equilibrium include improving societies, institutions and organisational practices, and modifying business and international relations.
We now present existing methods of motivating moving to the desired stable state.

When the players' utilities (preferences) are unknown, mechanism design~\cite[Chapter~$9$]{NisanRoughgardenTardosVazirani2007} and implementation theory~\cite{Jackson2001} devise a game form, such that for any profile of utility functions, the resulting game will have the desired equilibria.
For example, Grant et al.~\cite{GrantKrausWooldridgeZuckerman2014} studied
manipulating a Boolean game without Nash
equilibria by announcements, such that the manipulated game has a Nash equilibrium, satisfying some objective. 
The influence of taxation schemes on Nash equilibria appeared
in~\cite{EndrissKrausLangWooldridge2011}.

When the planner knows the players' utilities, Anshelevich and Sekar%
~\cite{AnshelevichSekar2014}
found a socially profitable strategy profile which was almost stable,
and proved that profile could be stabilised by a small
payment to the players. 
Monderer and Tennenholtz~\cite{MondererTennenholtz2004} subsidise certain outcomes to make 
a certain profile hold in non-dominated strategies, such that that profile
itself requires little, perhaps even
zero, subsidising. Allowing both taxing and subsidising, while requiring fairness, individual rationality, Polevoy and Dziubi\'nski~\cite{PolevoyDziubinski2022} made any desired profile strictly dominant by adjusting the payoffs
in a way that the total taxes were always equal to the total subsidies.

Given a game, one may communicate the focal point equilibrium~\cite{Schelling1980}, if the players are cooperative enough.
Our direct control sets can be seen as both motivating and communicating the equilibrium to play. 
The number of players implying a Nash equilibrium was studied by Kalai and Tauman~\cite{Kalai2020,TaumanKalai2023}, as we describe next.

Kalai~\cite{Kalai2020} studied the stability of Nash equilibria against deviations and of equilibrium formation by some players. He defined the \defined{defection index} of an equilibrium as the minimum number of deviators who can motivate some player to deviate, i.e.~no smaller coalition of deviators can motivate anyone to deviate.
Defection index had previously been studied by Abraham et al.~\cite{AbrahamDolevGonenHalpern2006} in the context of secret sharing.
Kalai also defined the \defined{formation index} as the minimum number of players such that any such subset forms that Nash equilibrium.
This line of work is theoretically justified by Tauman and Kalai~\cite{TaumanKalai2023}, who axiomatically characterise best-response equilibria of various viability and calculate the viability of some coordination games on graphs.
These indices, partially corroborated experimentally~\cite{KimMinWooders2022}, are monotonic, because the formation guarantee is required for \emph{any} coalition of a certain size. 

Unlike us, Kalai's indices refer to any non-compliant coalition of players rather than to a special coalition, thus being \defined{uniformly $m$-incentive compatible} in the sense of Tauman and Kalai~\cite{TaumanKalai2023} implies any coalition of size at least $m$ is a direct control set, but is unnecessary for that. Kalai did not tackle our algorithmic question of attaining an equilibrium. Other algorithmic questions, such as fault-tolerant computation, were studied by Ben-Or et al.~\cite{BenOrGoldwasserWigderson1988}, Eliaz~\cite{Eliaz2002}, who proposed \defined{$k$-fault-tolerant Nash equilibrium} which is equivalent to Nash equilibrium with a defection index greater than $k$,
and Abraham et al.~\cite{AbrahamDolevGonenHalpern2006}, who called defection index the \defined{level of resilience}.

Our approach of making the direct control set play as desired, thereby exerting influence on the others, can seem akin to creating a new Stackelberg game, where the control set players are the leaders. However, leaders in Stackelberg games optimise their own utilities, by considering the reaction of the followers, while the direct control set players do not optimise their own utilities.

%% file: gen_game.tex
\section{Control Sets vs Strong Nash Equilibria}\label{Sec:cntrl_vs_ne}

Since the group deviations are studied both in the definition of control sets and of strong Nash equilibria, we now study the connection between control sets and strong equilibria.
Recall
\begin{defin}
Given a game $(G, (s_i)_{i \in N}, (u_i)_{i \in N})$, a profile $s \in S$
is a \defined{k-strong \NE} if no coalition $K$ of at most $k$ players can
deviate, such that nobody loses and someone gains. In formulas,
$K \subseteq N, \abs{K} \leq k$ implies for each subprofile $d_K$ that 
$u_i(d_K, s_{N \setminus K}) \geq u_i(s), \forall i \in K
\Rightarrow \not \exists j \in K$, such that 
$u_j(d_K, s_{N \setminus K}) > u_i(s)$.
\end{defin}

Intuitively, having direct control sets implies nonsusceptibility to certain group deviations, thereby implying strength. 
Formally,
\begin{theorem}\label{the:cntl_set_str_eq}
Assume that game~$G$ satisfies both following conditions 
for some $d \in \NE$:
\begin{enumerate}
	\item Given any profile $s$,
	any subset of at least $m$ agents is a direct control set 
	with respect to $G, s$ and $d$.
	
	\item	For any player~$j$, playing $d_j$ in $d$ is at least
	as profitable for her as in any other profile. In formulas,
    $\forall j \in N, \forall s_{-j} \in S_{-j}, u_j(d) \geq u_j(d_j, s_{-j})$.
\end{enumerate}
Then, $d$ is an $n - m$-strong \NE,
and this is tight for every $m < n$.
\end{theorem}
We remark that the $2$nd assumption can be seen as stating 
that complying is most convenient when everyone else complies as well.
\begin{proof}
Consider any group deviation from $d$ by agents $A \subset N$,
which deviate to~$s_A$, where $\abs{A} \leq n - m$. 
By the $1$st assumption, the at least $n - (n - m) = m$ non-deviating 
agents constitute a direct control set with respect to $G, s$ and $d$.
Thus, for any $i \in A$, playing $d_i$ is a best response now,
in particular
$u_i(d_{N \setminus A}, s_{A \setminus \set{i}}, d_i) 
\geq u_i(d_{N \setminus A}, s_{A \setminus \set{i}}, s_i)$.

Now, assumption~$2$ implies that $\forall i \in A$, 
$u_i(d_{N \setminus \set{i}}, d_i) 
\geq u_i(d_{N \setminus A}, s_{A \setminus \set{i}}, d_i) 
\geq u_i(d_{N \setminus A}, s_{A \setminus \set{i}}, s_i)$.
Namely, no deviator enjoys the group deviation, implying 
the claimed strength of $d$.

The following example demonstrates tightness, that is for any $n$ and $m < n$, there exists an example satisfying the premises of the Theorem, where the $\NE~d$ is $n - m$-strong, but not stronger.
\begin{example}
Consider a symmetric game with $n$ players $N$, each player with the
strategy set $S = \set{a, b}$ and the following utility function for 
every player~$i$:
\begin{equation*}
\begin{cases}
	1 & \text{if } s_i = a \text{ and } \abs{\set{j \in N : s_i = s_j}} \geq m + 1, \\
	2 & \text{if } s_i = b \text{ and } \abs{\set{j \in N : s_i = s_j}} \geq n - m + 1, \\
	0 & \text{otherwise. } \\
\end{cases}
\end{equation*}
Here, profile $d \defas (a, a, \ldots, a)$ is
an \NE, because obtaining utility $2$ requires are least $n - m + 1 > 1$ players to play $b$. Given any $s \in S$, any subset of at least $m$ agents constitutes a direct control set w.r.t.~$G, s$ and $d$, because that leaves at most $n - m$ agents outside the direct control set. Assumption $2$ holds, too. As for the conclusion, $d$ is an $n - m$-strong \NE, but not stronger, because more than $n - m$ agents can deviate to $b$ and obtain $2$ each instead of $1$. This thus provides a tight example.
\end{example}
\end{proof}

In general, the converse does not hold, i.e.~$d$ being a $n - m$-strong \NE{}
does not generally imply assumption~$1$ above, under assumption~$2$. For example consider the
following $2$-player game.
\[ \begin{array}{l|c|c|}
			& L: & R: \\
\hline
T:	& 1, 1 & 0, 0 \\
\hline
B: 	& 0, 0 & 2, 2 \\
\hline
\end{array}\] 
Here, $d \defas (B, R)$ is a $2 = 2 - 0$-strong \NE. Additionally, assumption~$2$
holds. However, assumption~$1$ fails to hold, namely no $0$
agents constitute a direct control set with respect to this game,
$s \defas (T, L)$ and $d = (B, R)$.

\section{Hardness and Possible Solutions}

We now prove that the direct control set problem is immensely hard, even for approximation, and then cope with that hardness by considering subcases and approximations.

\subsection{Hardness}\label{Sec:hard}

The brute-force for ``casting'' all the subsets of increasing sizes
for the role of a control set can take exponential time. The question
arises whether the problem is intrinsically hard. Indeed,
we now prove the hardness of the decision version of the
Direct control set problem (Definition~\ref{def:dir_control_set_prob}).

\begin{theorem}
The decision version of the direct control set
problem is \NP-complete.
\end{theorem}
\begin{proof}
The \NP-hardness holds even for the restricted case of coordination
games on graphs, as we will show in Theorem~\ref{The:dec_cont_set_npc}.

These problems belong to \NP, the witness being a direct control set.
Since verifying a set of agents being a control set
can simply be done
by making sure that destination equilibrium strategies are optimal (best response)
for each agent outside the control set, this problem belongs to \NP.
\end{proof}

Decision hardness generally does not preclude finding a direct control set that is close to the minimum possible size. However, even this is very hard, as we now prove.
\begin{theorem}
The problem of finding a direct control set of the minimum size is not approximable 
within factor~$n^{1 - \epsilon}$, for any $\epsilon > 0$, unless $\Pol = \NP$.
This holds even when we restrict the direct control set problem to
bring an \NE{} to an \NE{}, in a game where each player has $2$ strategies.
\end{theorem}
\begin{proof}
Consider the problem of deleting a minimum vertex set from a connected graph $G = (V, E)$ to obtain a tree, namely a connected acyclic subgraph. This problem is not approximable within factor~$\abs{V}^{1 - \epsilon}$, for any $\epsilon > 0$, unless $\Pol = \NP$~\cite{Yannakakis1979}.

Given an instance of the problem of deleting a minimum vertex set from a connected graph $G = (V, E)$ to obtain a tree, we define a game 
with players $V$ and actions $S_1 = \ldots = S_n = \set{a, b}$ each.
Let the utilities of every player~$l \in N$ be 
\begin{equation*}
u_l(s) \defas
\begin{cases}
	2 & \text{The subgraph induced by } V \setminus \set{i \in V \setminus\set{l} : s_i = b} \text{ is a tree}, \\
    1 & \text{otherwise, if } s_i = a, \\
    0 & \text{otherwise.}
\end{cases}
\end{equation*}
Here, profile $s \defas (a, a, \ldots, a)$ is an \NE, because the empty set is not a feasible solution, because we can assume w.l.o.g.\ that the given instance is not a tree, since that can be checked in polynomial time and such instances can be reduced to a predefined instance of the direct control set problem. Profile $d \defas (b, b, \ldots, b)$ constitutes an \NE, because deleting all the nodes besides one always yields a tree. Finally, a set of vertices is a direct control set w.r.t.~$G, (a, \ldots, a), (b, \ldots, b)$ 
if and only if it solves the given problem of deleting a vertex set to obtain a tree. The above hardness results implies the theorem.
\end{proof}

Having removed all hope for a decent polynomial approximation of the general problem, we cope with that by 
providing a non-polynomial algorithm in \sectnref{Sec:Inc_alg}, by solving the case where the graph of the game is a tree using dynamic programming in \sectnref{Sec:DP_tree}, and finally by approximating interesting subclasses in \sectnref{Sec:mono_alg}, assuming monotonicity with respect to inclusion that we define there.

\subsection{Incremental Algorithm}\label{Sec:Inc_alg}

We first present the incremental Algorithm~\ref{alg:perm_add}, adding players till we obtain a direct control set, in all the possible orders.
This algorithm terminates with a correct result, 
because it tries all the possible orders. 
\begin{algorithm}[ht!]
\caption{\textbf{MinDCS} $\paren{G = (N, S, (u_i)_{i = 1, \ldots, n}), s \in S, d \in \NE, w \colon N \to \realsP}$}\label{alg:perm_add}
\SetAlgoVlined
\LinesNumbered

         $\calS \leftarrow \emptyset$;

        \For{any ordering $R$ of the players}{
            $S \leftarrow \emptyset$;\\
            \While{$S$ is NOT a DCS}{
                $S \leftarrow S \cup \set{i}$; \tcp{Add players one by one in that order}
            }
            $\calS \leftarrow \calS \cup S$. \tcp{Append $S$, if it's not there yet}
        }
        \Return{A minimum-weight set among $S \in \calS$. }

\end{algorithm}
This process generally
requires exponential time 
(trying all the permutations and checking being a control set) 
and space (for keeping all the found control sets). 
However, if many addition orders quickly reach control sets, 
this process will terminate speedily and use up less memory
as well. 

\subsection{Dynamic Programming on Trees}\label{Sec:DP_tree}

We now describe an algorithm that solves the problem using dynamic programming, when the dependencies of the players' utilities 
are representable by a tree; in other words, when we have a tree graphical game. More precisely, given a strategic game 
$G = (N, S = S_1 \times S_2 \times \ldots \times S_n, (u_i)_{i = 1, \ldots, n})$, one defines the graph of this game~\cite{KearnsLittmanSingh2001} as the undirected graph on the set of the players $(N, E)$, where a player's utility can depend on her neighbours only. Formally, there exists a matrix $U_i$ for every player~$i \in N$, such that her utility is $u_i(s) = U_i(s_{N(i)})$, where $N(i) \defas \set{j \in N : \exists (i, j) \in E} \cup \set{i}$.
If this graph is a tree, then Algorithm~\ref{alg:DP_tree} solves the problem. The Boolean parameter $par$ tells whether the parent in the tree is chosen to the direct control set; it is set arbitrarily when the vertex~$i$ we are dealing with is the root.

\begin{algorithm}[t!]
\caption{\textbf{MinDCS\_Tree} $\paren{G = (N, S, E, (U_i)_{i = 1}^n), s \in S, d \in \NE, w \colon N \to \realsP, par, i}$}\label{alg:DP_tree}
\SetAlgoVlined
\LinesNumbered

        \If {$i$ is a leaf}{
            \If {$\emptyset$ constitutes a DCS for $i$ and her parent}{
                \Return{$\emptyset$};
            }
            \Else{
                \Return{$\set{i}$};
            }
        }
        \Else {  \tcp{the recursion}
            
            $DCS_2 \leftarrow \emptyset$ \tcp{Take $i$ to the $DCS_2$}
            \ForEach{$c \in Ch$}{
                $DCS_2 \leftarrow DCS_2
                \cup \textbf{MinDCS\_Tree} \paren{G = (N, S, E, (U_i)_{i = 1}^n), s \in S, d \in \NE, w \colon N \to \realsP, true, c}$
                }
            \Return{$\argmin_{X \in \calC}{\sum_{x \in X}w(x)}$};
        }
\end{algorithm}

Algorithm~\ref{alg:DP_tree} goes over the tree recursively, and tests whether a node should be taken to the direct control set, based on whether it is necessary for its parent and itself, and if it is optional, on what will happen in the subtrees.

\subsection{Monotonicity-Based Algorithms}\label{Sec:mono_alg}

We now provide 
algorithms
for finding a direct control set in the cases that are monotonic with respect
to inclusion, as one may intuitively hope. 
\begin{defin}\label{def:mono}
\defined{Monotonicity with respect 
to inclusion} means that if $A$ is a direct control set with respect
to game $G$ and solutions $s$ and $d$, so is any including
set $A' \supset A$. 
\end{defin}
In other words, being a direct control set
cannot be vitiated by adding players.

Despite the intuitive appeal of monotonicity, it does not have to hold
even in the simple case of singleton congestion games, as the following example demonstrates.
\begin{example}\label{ex:no_mono_sing_cong}
Consider the singleton congestion game with $k$ resources $R = \set{0, 1, \ldots, m - 1}$ and players $N = \set{1, 2, \ldots, n}$, where $m$ divides $n$. Let each resource's cost function be $c_r(x) \defas x,  \forall r \in R$. Assume each player chooses a resource, namely each player $i$'s strategy set is $S_i \defas R$.

Now, let profile $s$ be the profile where the first $n / m$ players choose resource $0$, the next $n / m$ players choose resource $1$, and so on. This constitutes an \NE. Consider the profile obtained by shifting each player to the next resource, cyclically wrapping around the end; formally, if $s_i = \set{r}$, then let $d_i$ be $\set{r + 1} \mod m$. This is also an \NE. The below illustration demonstrates the players (lower row) choosing the resources (upper row) in profile~$s$ (solid lines) and in~$d$ (dashed lines).

\begin{tikzpicture}[
  node distance=1.5cm,
  vertex/.style={circle, draw, fill=white, inner sep=0pt, minimum size=15pt}
]

 Original correspondence
\foreach \i in {0,1,...,2} {
 \node[vertex] (res\i) at (2 * \i, 1) {};
  \foreach \j in {0,1} {  
    \node[vertex] (player\i\j) at (2 * \i + \j, 0) {};
    \draw (res\i) -- (player\i\j);
  }
}

\foreach \i [evaluate=\i as \nextres using {int(mod(\i + 1, 3))}] in {0,1,...,2} {
  \foreach \j in {0,1} {
    \draw[dashed] (player\i\j) -- (res\nextres);
  }
}


\end{tikzpicture}

We now demonstrate that the direct control set problem of bringing $s$ to $d$ is not monotonic. First, denote all the players playing $\set{r}$ in profile $x$ as $P_r(x)$, namely $P_r(x) \defas \set{i \in N : x_i = \set{r}}$. Now, we notice that the set $\cup_{r \in R : r = 2k, k \in \naturals}{P_r(s)}$ is a direct control set, because after those players switch to their strategies in $d$, any other player will be motivated to play her strategy in $d$, too. However, adding any other player, say a player from~$P_1(s)$, to the above directed control set, renders that set stop being a direct control set, because moving to resource~$2$ will not be a best response for the other players from~$P_1(s)$ anymore.
\end{example}

For the sake of algorithms that we present in this section,
we need to look at the structure of the control sets in terms of the
agents required to control a given agent's best response, rather than only
at the set controlling everyone's best response. Then, we will define 
monotonicity of such sets as well.
\begin{defin}
Consider any agent $k \in N$.
Given solutions $s$ and $d$, if either $k \in A$, or the agents $A \subseteq (N \setminus \set{k})$ can play
strategies $(d_i)_{i \in A}$ such that agent $k$'s best response to
$(d_i)_{i \in A}, (s_j)_{j \in N \setminus A}$ is $d_k$, then we say $A$
can \defined{(directly) bring $k$ from $s$ to $d$}. \\
Such a set $A$ is called a \defined{direct control set for $k$} with respect to $G$,
$s$ and $d$. \\
\end{defin}

We immediately observe the following.
\begin{observation}\label{obs:dir_cntl_set_player_all}
Set $A \subseteq N$ is a direct control set with respect to
$G$, $s$ and $d$
if and only if for each $k \in N \setminus A$, the agents $A$
constitute a direct control set for $k$ with respect to
$G$, $s$ and $d$.
\end{observation}
%

We now define monotonicity that holds per player, in addition to the monotonicity of Definition~\ref{def:mono}, namely
\begin{defin}
\defined{Player-wise monotonicity with respect 
to inclusion} means that for any $k \in N$, if $A$ is a direct control set for $k$ with respect
to game $G$ and solutions $s$ and $d$, so is any including
set $A' \supset A$.
\end{defin}
By Observation~\ref{obs:dir_cntl_set_player_all}, player-wise monotonicity implies Definition~\ref{def:mono}. Therefore, Example~\ref{ex:no_mono_sing_cong} violates also player-wise monotonicity.
The converse implication does not generally hold, because $A$ may be a direct control set for some, but not all players, therefore not being a direct control set with respect to $G, s$ and $d$. 

In this subsection, we assume player-wise monotonicity w.r.t.~inclusion 
by default, and provide approximation algorithms under that assumption.

\subsubsection{Reductions}

We first provide reductions between our problem and Hitting set, and then move to concrete algorithms for our problem.
Assume that for each player, every minimal direct control set%
\footnote{Minimal direct control set means a direct control set which any proper subset is not a direct control set.}
is a singleton (namely, any control set that contains more that one 
player contains a proper subset that is a control set, too). Then,
the problem can be distilled to finding a hitting set, as we now prove.

\begin{proposition}
Assume that in any game $G = (N, (S_i)_{i \in N}, (u_i)_{i \in N})$, for any profile $s \in S$ and equilibrium $d \in \NE$, and for every player~$i$, every minimal direct control set is a singleton, namely if $D \subset N$ is a direct control set of size greater than $1$, then there exists a subset $D' \subsetneq D$, which is also a direct control set. In such cases, the approximation hardness of the direct control set problem is equivalent to the hitting set problem~\cite{Karp1972}, which can be approximated~\cite{Johnson1974}.
\end{proposition}
\begin{proof}
Given a direct control set problem ($G, s, d$), we reduce it to hitting set as follows:
\begin{enumerate}
    \item For each $i \in N$, exhaustively check which singletons constitute direct control sets for that player. Denote the union of all such singletons for player~$i$ by $DCS(i)$. (Player~$i$ always belongs to $DCS(i)$.)
    \item Solve the hitting set for the collection $(DCS(i))_{i \in N}$, and return the obtained set.
\end{enumerate}
Indeed, a set is direct control set if and only if it contains at least one singleton direct control set, which is exactly what a hitting set of the collection $(DCS(i))_{i \in N}$ means.

As for hardness, we provide a reduction from hitting set to direct control set with singleton minimum direct control sets. Given a hitting set problem, namely a collection $(E_l)_{l \in I}$ of subsets of ground set $U$, we construct a game on players $N \defas I \dot\cup (\cup_{i \in I}{E_l})$.
The weight of every player from $I$ is any large enough $M$, to prevent choosing those to direct control sets, while any other player weighs $1$.
Each player has a binary choice, namely $S_i \defas \set{0, 1}$, and
we define the utility, such that each player $l$ in $I$ will be controlled by any of the players $E_l$. Formally, the utility of each player outside $I$ is some constant, while for each $l \in I$, we define
\begin{equation*}
u_l(s) \defas 
\begin{cases}
	1 & \text{if } s_l = 1 \text{ and } \exists i \in E_l, s_i = 1, \\
    -1 & \text{if } s_l = 1 \text{ and } \not\exists i \in E_l, s_i = 1, \\
    0 & \text{otherwise}. \\
\end{cases}
\end{equation*}
The goal of the constructed hitting set problem is to move from the \NE{} where every player plays $0$ to the equilibrium where everyone chooses $1$.

Here, the solutions to the given hitting set problem are mapped on the solutions to the constructed direct control set problem with the same cost. This mapping is one-to-one, and every direct control set can result from it, besides sets with at least one element of weight $M$.
\end{proof}
The above proof also applies to reducing weighted hitting set to weighted  direct control set and the other way around.

Another interesting case occurs when for each player, we are given
all the minimal 
direct control sets. 
This generalises the case above, thus it is at least as hard
as hitting set. This is actually an instance of
the Minimum Hitting Set of Bundles Problem, where at least one
bundle of each set of bundles has to be 
covered~\cite{AngelBampisGourvès2009,Damaschke2015}. 
We can also use the approximation results for 
the Minimum Hitting Set of Bundles Problem by reducing conversely,
similarly to how we reduced the above case to hitting set.

\subsubsection{Local Ratio}

When the number of players whose strategies influence players $i$'s utility is bounded by a (perhaps constant) function $f$, for each $i \in N$, we can provide
a \defined{Local Ratio} approximation. The local ratio technique~\cite{BarYehudaEven1985,BarNoyBarYehudaFreundNaorShieber2001,BarYehudaBendelFreundRawitz2004} is based on finding a weight function such that every
feasible solution is an $r$-approximation. Then, subtracting this function and solving the resulting problem recursively, which of course can be also implemented iteratively, we inductively obtain an $r$-approximation for the resulting problem. Then, that solution is also an $r$-approximation w.r.t.~the original weight function, by the 
\begin{theorem}[{Local Ratio Theorem~\cite[Theorem~$2$]{BarYehudaBendelFreundRawitz2004}}]
\label{The:LR_thm}
Let weight functions $w, w_1, w_2$ satisfy $w = w_1 + w_2$, and let any feasible solution satisfy $x$ be an $r$-approximation w.r.t.~both $w_1$ and to $w_2$.
Then, $x$ is also an $r$-approximation w.r.t.~$w$.
\end{theorem}

Consider the Local Ratio Algorithm~\ref{alg:LR_int}, which takes any not controlled player~$i$ and defines an equal weight for each of the players whose strategies sometimes influence $i$, and zero weight for every other player. Subtracting this weight, the algorithm recursively invokes itself and returns its output. Since any feasible direct control set contains at least one of the influencing players and at most all of them,
any feasible solution is a ``number of influencing players''-approximation.
%
\begin{algorithm}[t!]
\caption{\textbf{MinDCS} $\paren{G = (N, S, (u_i)_{i = 1}^n), s \in S, d \in \NE, w \colon N \to \realsP}$}\label{alg:LR_int}
\SetAlgoVlined
\LinesNumbered

         $S \leftarrow \set{i \in N : w(i) = 0}$;
        
        \If {$S$ constitutes a DCS}{
            \Return{$S$}.
        }
        \Else{
            Take any player~$i \in N \setminus S$, s.t.~$d_i \not \in BR_i(d_S, s_{N \setminus S})$; \tcp{Take any not yet controlled player}
            Let $N(i) \leftarrow \set{j \in N \setminus S : i \text{ can influence } u_i}$;\\
            \tcp{$i$ is NOT controlled yet $\Rightarrow$ at least one and at most all of $N(i)$ should be in any feasible DCS}
            Set $\epsilon \leftarrow \min_{j \in N(i)}{w(j)}$;\\
            Let $w_2 \defas 
            \begin{cases}
                \epsilon & \text{if } j \in N(i), \\
                0 & \text{otherwise; } \\
            \end{cases}$
            \tcp{W.r.t.~$w_2$, every feasible DCS is an $\abs{N(i)}$-approximation}
            \Return{
                \textbf{MinDCS} $\paren{G = (N, S, (u_i)_{i = 1}^n), s \in S, d \in \NE, w - w_2}$.\label{alg:LR_int:rec}
            }
        }
\end{algorithm}

Algorithm~\ref{alg:LR_int} terminates in polynomial time, since it performs actions taking polynomial time each, such as checking feasibility of a DCS, finding the influencing players and defining the new weight function, at most $n$ times.
Finally, we formally bound the approximation ratio of Algorithm~\ref{alg:LR_int}.
\begin{theorem}
Algorithm~\ref{alg:LR_int} returns a feasible $f$-approximate direct control set.
\end{theorem}
\begin{proof}
We prove the $f$-approximation with respect to $w$ by induction on the recursion of the algorithm.

In the induction basis, the algorithm returns a feasible zero-weight, and thus optimum, solution.

For an induction step, assume we are in the ``else'' branch, namely $S$ is not a direct control set, Then, there exists some player $i$, whose best response to $d_S, s_{N \setminus S}$ does not include $d_i$. Consequently, any feasible direct control set must contain at least one player who can influence $i$'s utility, namely a member of $N(i)$, therefore weighing at least $\epsilon$. Since any set contains at most $\abs{N(i)} \leq f$ players of $N(i)$, any feasible direct control set is an $f$-approximation w.r.t.~$w_2$. On the other hand, the induction hypothesis implies that the invocation in line~\ref{alg:LR_int:rec} returns a feasible $w - w_2$-approximation. Finally, the Local Ration Theorem~\ref{The:LR_thm} implies the returned solution is also a $w_2 + (w - w_2) = w$-approximation. The feasibility stems from the induction hypothesis.
\end{proof}

Naturally, this algorithm is only applicable when no player's utility can be influenced by many players, i.e.\ for a small~$f$. For example, this holds in congestion games where every player shares resources with only few other players.

\section{Potential Games}\label{Sec:pot}
Potential games~\cite{MondererShapley1996PG}
are defined as the set of strategic form games $G = (N, S = S_1 \times S_2 \times \ldots \times S_n, (u_i)_{i = 1, \ldots, n})$ such that there exists a function $P : S \to \reals$, with the property that for any subprofile $s_{-i} \in \times_{i \in N \setminus \set{i}}$ 
and for any two strategies $s_i, t_i$ of player~$i$, her utility difference when moving from $s_i$ to $t_i$ is equal to that of $P$, namely
$u_i(s_i, s_{-i}) - u_i(t_i, s_{-i}) = P(s_i, s_{-i}) - P(t_i, s_{-i})$.
Within the study of switching between stable equilibria, potential games are naturally interesting to study since they always have a Nash Equilibrium. Also, Monderer and Shapley~\cite{MondererShapley1996PG} proved that the subset of finite potential games is isomorphic to the set of congestion games, 
which is the set of strategic games $(N, R, (S_i )_{i \in N }, (c_j )_{j \in R})$ where $R$ is a set of resources used by the strategies and $c_j$ are cost functions to be paid by the players for using those resources. 

In this section, 
we show that in the case of singleton congestion games i.e. the subset of such games where each strategy uses exactly one resource, this bound can be reduced to $\lfloor n-n/m\rfloor$ where $m$ is the number of resources in $G$. Additionally, we present a polynomial time algorithm which, under some additional assumptions, finds an optimal direct control set for a given instance $(G,s,d)$ for a singleton congestion game $G$. 
We then analyse further subclasses of potential games, namely symmetric games in~\sectnref{Sec:pot:symm}
and coordination games on graphs in~\sectnref{Sec:coord_game_graph}.




\subsection{Singleton Congestion Games}\label{Sec:pot:single_cong}
Our analysis of singleton congestion games heavily uses two core concepts. The first is the strategy profile that, given $(G,s,d)$ and a DCS $A$, results when we begin with the initial profile $s$ and switch the strategies of all players in $A$ to their strategy in $d$.
We call the profile that results from these changes the \emph{intermediate profile} with respect to $(G,s,d)$ and $A$ and denote it by $sd(A)$. 
By the definition of direct control sets, it is a necessary property of an intermediate profile $sd(A)$ that for all players $i \in N\setminus A: d_i \in Br_i(sd(A))$. Further, if we consider any strategy profile $s$ in a singleton congestion game, it is easy to see that for any pair of players for whom it is a best response to switch strategies in $sd(A)$, the set of possible alternatives is shared. Formally, for any players $i \neq j$ with $\emptyset \neq Br_i(sd(A)) \neq sd_i(A)$ ($j$ respectively) it holds that $Br_i(sd(A)) \setminus sd_i(A) = Br_j(sd(A)) \setminus sd_j(A)$. This leads to the following definition:

\begin{defin}
    Given a strategy profile $s$ of a singleton congestion game $(N, R, (S_i )_{i \in N }, (c_j )_{j \in R})$ we call the set $AB(s) = \bigcup_{i\in N} (Br_i(s) \setminus s_i)$ the \defined{attraction basin} of $s$.
\end{defin}

While formally, the attraction basin is a set of strategies, we often refer to it as a set of resources in the context of singleton congestion games since in this case, each strategy represents exactly one resource. We immediately observe some properties of attraction basins summarized in Observation \ref{obs:attract_basin_props}.

\begin{observation}
    For any attraction basin $AB(s)$ it holds:
    \begin{itemize}
        \item For all $q,r \in AB(s)$ we have $, c_q(|P_q(s)|+1) = c_r(|P_r(s)|+1)$
    \end{itemize}
    \label{obs:attract_basin_props}
\end{observation}
The first holds since, otherwise, one of the resources in AB(s) would be a better best response than another, which is a contradiction to the definition of attraction basins as a union of best responses.

\begin{lemma}
    For any singleton congestion game $G$, a minimal DCS $A$ with respect to profiles $s,d$, has at most size $\lfloor n-n/m\rfloor$ and this bound is tight. 
\end{lemma}
\begin{proof}
    We first show that such a DCS always has to exist, which implies that any larger DCS is not minimal. And then show the tightness of the bound by presenting an examplary (G,s,d) with minimal DCS of exactly this size.

    We can always construct a DCS of the claimed size in the following way. First, choose a single resource $r$ that maximizes $|P_r(d)|$. Then take into the DCS all players $A_1 = \bigcup_{q \neq r} P_q(d)$. Thus, the only players not in $A_1$ are those that are supposed to play $r$ in $d$. Since $d$ is assumed to be a NE, it follows that playing $r$ in the intermediate profile is a best response for all players that are supposed to play $r$ in $d$, which implies $AB(sd(A_1) = r$. And since all other players are in $A_1$, it follows that it indeed is a DCS. By construction, we have $|A_1| = |N \setminus P_r(d)|$. Since we chose $r$ to maximize $|P_r(d)|$, it can be at worst $\lceil n/m\rceil$ which happens when in $d$ all players are distributed as equally as possible among the resources. Thus, we have $|A_1| \leq \lfloor n - n/m\rfloor$.

    For the tightness, consider a singleton congestion game $G$ where the cost function for each resource $j$ is defined as $c_j(x) = j+m*(x-1)$. Now we define $(G,s,d)$ such that for each player $i$ we have $s_i \neq d_i$. Thus all players are supposed to move. Note that the cost functions have no equal integer values, which implies that $|AB(sd(A_2))| \leq 1$ for every possible intermediate profile of any DCS $A_2$. Thus, it is impossible to make more than one resource $r$ a best response at the same time. Since every player is supposed to change strategies from $s$ to $d$, all players in $N \setminus P_r(d)$ need to be moved by being taken into the DCS, which results in the claimed bound.
\end{proof}

In this proof, we can additionally observe that the set of possible attraction basins in a singleton congestion game is directly linked to the set of possible direct control sets.
Therefore, in the next part, we tackle the problem of finding a minimal direct control set by searching for a set of resources that can be made the attraction basin in an intermediate profile while needing the least number of players in the DCS.
As a first step, we show in Observation \ref{obs:SingletonAB_Structure} some immediate properties induced by considering a fixed set of resources as the attraction basin. Note that these of course only hold if it is at all possible to make the chosen set of resources an attraction basin by only moving players from $s$ to $d$.

\begin{observation}
\label{obs:SingletonAB_Structure}
    In a singleton congestion game $G$ and profiles $s,d$, for a given intermediate profile $sd(A)$ which induces an attraction basin $AB(sd(A))$ the direct control set $A$ has the following properties:
    \begin{align}
        \{P_r(d) \cap P_q(s) : q,r \notin AB(sd(A))\}  &\in A\\
        \{P_r(d) \cap P_q(s), : r \notin AB(sd(A)), q \in AB(sd(A))\} &\in A\\
        \{i \in P_r(d) \cap P_r(s) : r \notin AB(sd(A)), Br_i(sd(A) = AB(sd(A)) \} &\in A
    \end{align}
\end{observation}

These properties directly follow from the definition of attraction basins and direct control sets and are necessary for every DCS in singleton congestion games. For the rest of this section, we use the union of these sets as a basis in order to construct a minimal DCS given that the set of resources that form the attraction basin is fixed. We then use the result to establish an algorithmic solution to the problem of finding a minimal DCS by computing a DCS for various possible attraction basins and choosing the smallest one.
A difficulty in this approach is the number of subsets of resources that can form an attraction basin, which can be exponential. However, in real-world applications, it is unlikely that the cost functions of all resources can be equalized within the parameter interval $[1, \dots, n]$. It is reasonable to assume that the number of subsets that can be equalized is limited, especially since the parameters are integers. Therefore, we make the following assumption:
\begin{assumption}[General Position]
    In a singleton congestion game, for each profile  $s$ it holds that $|AB(s)| \leq 1$.
\end{assumption}
With this assumption, the number of attraction basins we need to consider becomes linear in the number of players.
We observe:

\begin{observation}
\label{obs:singletonMinusDCSStructure}
    If we assume $|AB(sd(A))| \leq 1$, the union of the sets in Observation \ref{obs:SingletonAB_Structure} include all players with the exclusion of three sets:
    \begin{align}
        \{ P_r(d) \cap P_r(s) &: r \in AB(sd(A))\} \label{align:singleton:stayInAB} \\
        \{ P_r(d) \cap P_q(s) &: r \in AB(sd(A)), q \notin AB(sd(A))\} \label{align:singleton:MoveToAB}\\
        \{i \in P_r(d) \cap P_r(s) &: r \notin AB(sd(A)), d_i \in Br_i(sd(A)) \label{align:singleton:stayOutsideAB} \} 
    \end{align}
\end{observation}
    
Taking any player from (\ref{align:singleton:stayInAB}) into the DCS would not change the intermediate profile; the only change would be a needless increase in DCS size. Thus, they are never part of a minimal DCS. Intuitively, similar arguments could be made for the other two sets since moving a player to AB who already wants to go there seems unnecessary. But, there is a possible trade-off between (\ref{align:singleton:MoveToAB}) and (\ref{align:singleton:stayOutsideAB}). 
It is possible to increase the size of (\ref{align:singleton:stayOutsideAB}) by taking players from (\ref{align:singleton:MoveToAB}) into the DCS until the cost of moving to AB is high enough that for at least some players in $\{i \in P_r(d) \cap P_r(s) : r \notin AB(sd(A)), Br_i(sd(A) = AB(sd(A)) \}$ their best response changes such that $d_i \ in Br_i(AB(sd(A))$. 
Generally, this trade-off can have an exponential number of solutions. Thus, we need to limit the limit the set we consider. To do this, we so far rely on the additional assumption that 
\begin{align}
    C_q(l_q(d)+1)>C_r(l_r(d)) \Rightarrow C_q(l_q(d))>C_r(l_r(d)-1) \label{align:singleton:AdditionalAssumption}
\end{align} 
holds for all pairs $r \neq q \in R$. Intuitively, this implies a certain linearity in the cost functions of the resources but is limited to only their values in the narrow scope of their loads in $d$. While this may not always be true, it is not unreasonable and at least easy to check in linear time.
We will now show that under this assumption, the aforementioned trade-off can only change the best response of players in $\{i \in P_r(d) \cap P_r(s) : r \notin AB(sd(A)), Br_i(sd(A) = AB(sd(A)) \}$ if at most one player is missing in $r$. Thus, the number of solutions is reduced from exponential to linear in the number of players.

If we assume \ref{align:singleton:AdditionalAssumption} and consider an arbitrary pair of resources $r \neq q$, we know that $C_q(l_q(d)+1) \geq C_r(l_r(d))$ holds since $d$ is a NE. If the inequality holds strictly, we immediately get $C_q(l_q(d))>C_r(l_r(d)-1)$ which implies that $r$ is a strict best response for players in $q$ if $l_(s) < l_r(d)-1$ for any possible intermediate profile $s$. If otherwise $C_q(l_q(d)+1) = C_r(l_r(d))$ holds, then in profile $d$, $q$ is a best response for players in $r$. Thus, we get $q \in AB(d)$, which by the general position implies $\set{q} = AB(d)$. In this case, we have that for all players $i \in P_r(d)\setminus P_r(s)$ also $i \in P_q(s)$ must hold, since otherwise we could construct an auxiliary DCS $B =N\setminus \set{i}$ where all players aside $i$ play $d$. If, in this case, $i \notin P_q(s)$ would hold, it would imply $\set{r,q} \subseteq AB(B)$, which would contradict the 
general position assumption. Thus, to prevent players $j \in P_q(d) \cap P_q(s)$ from having $r$ as a sole best response, it is necessary to move players to $r$ until at least $C_q(x) = C_r(y)$ is achieved, 
which we know is the case when $C_q(l_q(d)+1) = C_r(l_r(d))$ which again means that we need to move all but at most one player to $r$ to make the trade-off. Thus, we get the following Lemma:

\begin{lemma}
    \label{Lemma:singletonAB_Structure2}
    Consider a singleton congestion game $G$ and profiles $s,d$, where an attraction basin $AB$ is always at most a singleton $\forall s \in S: |AB(s)| \leq 1$. If we assume $C_q(l_q(d)+1)>C_r(l_r(d)) \Rightarrow C_q(l_q(d))>C_r(l_r(d)-1)$, then for a given intermediate profile $sd(A)$ which induces an attraction basin $AB(sd(A))$ a minimal direct control set $A$ has the following structure
    \begin{align}
        A   & = \{P_r(d) \cap P_q(s) : q,r \notin AB(sd(A))\} \\
            & \cup \{P_r(d) \cap P_q(s), : r \notin AB(sd(A)), q \in AB(sd(A))\} \\
            & \cup \{i \in P_r(d) \cap P_r(s) : r \notin AB(sd(A)), Br_i(sd(A) = AB(sd(A)) \} \setminus (\ref{align:tradeOff}) \\
            & \cup \set{
            {\text{Either: All, one, or none from}}\atop
            {\{ P_r(d) \cap P_q(s) : r \in AB(sd(A)), q \notin AB(sd(A))\}  } 
            } \label{align:tradeOff}      
    \end{align}
\end{lemma}

Thus, we can find a minimal DCS for a given congestion game algorithmically by iterating over all $AB$ and creating the set of all sets that given this $AB$ has the structure indicated by Lemma \ref{Lemma:singletonAB_Structure2}. A pseudocode is given in the following algorithm:

\begin{algorithm}[ht!]
\label{algorithm:singletonCongestion}
\caption{\textbf{SingletonMinDCS} $\paren{G = (N, S, R, (c_j)_{j = 1, \ldots, m}), s \in S, d \in \NE}$}\label{alg:singleton}
\SetAlgoVlined
\LinesNumbered

         $\calS \leftarrow \emptyset$ (set of min-DCS candidates);\\
         $\calS \leftarrow \calS \cup \bigcup_{r \in R} P_r(d)\setminus P_r(s)$ ("Full" DCS)\\
        \For{each $r \subset R$}{
                $DCS_\emptyset \leftarrow \set{P_p(d) \cap P_q(s) : p,q \neq r} \cup \set{ P_q(d) \cap P_r(s) : q\neq r}$\\
            \For{each $X \in \set{\set{i}: i \in p_r(d) \setminus P_r(s)}\cup \set{N}$}{
                $DCS_X \leftarrow (P_r(d)\setminus (P_r(s)\cup X)) \cup DCS_\emptyset$ \\
                $M \leftarrow \set{i \in P_q(s)\cap P_q(d): q \neq r, Br_i(sd(DCS_X = \set{r}}$\\
                $DCS_X \leftarrow DCS_X \cup M$\\
                $\calS \leftarrow \calS \cup DCS_X$\\
            }
        }
        \Return{A minimally sized set among $S \in \calS$. }
\end{algorithm}

 Under the assumption of general position and \ref{align:singleton:AdditionalAssumption}, the correctness of this algorithm follows from the correctness of Lemma \ref{Lemma:singletonAB_Structure2}. The algorithm's runtime is polynomial in the input size since the number of attraction basins is linear in the number of resources, and the number of DCS candidates per $AB$ is linear in the number of players. Thus, we get the following Lemma:

 \begin{lemma}
     Given $(G,s,d)$ with a singleton congestion game $G$ and Nash-equilibria $s,d$, there is a polynomial time algorithm that finds a minimal DCS with respect to $G,s,d$ if for all profiles $t$ $|AB(t)|\leq 1$ and $C_q(l_q(d)+1)>C_r(l_r(d)) \Rightarrow C_q(l_q(d))>C_r(l_r(d)-1)$ holds.
 \end{lemma}

\subsection{Symmetric Congestion Games with Decreasing Costs}\label{Sec:pot:symm}

Given a symmetric congestion game with decreasing costs,
consider bringing an \NE{}~$s$ to \NE~$d$. We will prove that
in any Nash equilibrium, all the players play the same strategy,
and therefore, only
the size of the control set matters. This immediately allows for
a brute force algorithm for direct control set. 

We consider a congestion game
$G = (N, R, S_1 \subseteq 2^R, \ldots, S_n \subseteq 2^R, (c_r)_{r \in R})$
and denote the total cost player~$i$ experiences by $C_i(x) \defas \sum_{r \in x}{c_r(l_r(x))}$.
\begin{proposition}
Consider a symmetric congestion game with decreasing costs
$G = (N, R, S_1 = \ldots = S_n \subseteq 2^R, (c_r)_{r \in R})$,
where $R$ are the resources, $S_1 = \ldots = S_n$ are the common strategy
sets and $(c_r)_{r \in R}$ are the decreasing costs.
Then, in any Nash equilibrium $d$, all the players play identical
strategies, namely $d_1 = \ldots = d_n$.
\end{proposition}
\begin{proof}
Assume by contradiction that a Nash equilibrium~$d$ has players playing different
strategies, w.l.o.g., $d_1 \neq d_2$. Since $d \in \NE$,
$C_1(d) \leq C_1(d_2, d_2, d_{-\set{1, 2}})$. Because the costs are decreasing,
$C_2(d_1, d_1, d_{-\set{1, 2}}) < C_1(d)$, and therefore
$C_2(d_1, d_1, d_{-\set{1, 2}}) < C_1(d) \leq C_1(d_2, d_2, d_{-\set{1, 2}})
= C_2(d_2, d_2, d_{-\set{1, 2}}) < C_2(d_1, d_2, d_{-\set{1, 2}}) = C_2(d)$,
the last inequality stemming from the assumption that the costs decrease.
This means that player~$2$ has a profitable deviation to $d_1$, contradictory
to $d$ being a Nash equilibrium.
\end{proof}

The Proposition implies that when we want to bring $s \in \NE$ to $d \in \NE$ directly,
we only care about the number of players in the direct control set. Therefore,
we can trivially compute the smallest size of direct control set that suffices.

%% file: coord_game_graph.tex
\subsection{Coordination Games}\label{Sec:coord_game_graph}

This section studies coordination games (on graphs), inspired by
and generalising~\cite{AptdeKeijzerRahnSchaferSimon2017}.
\begin{defin}
A \defined{coordination game}~\cite{AptdeKeijzerRahnSchaferSimon2017} is
defined by an undirected graph $G = (N, E)$ without self-loops, which nodes are the
set of players. Each player~$i$ has a set of strategies (colours) $S_i$,
and the utility of player $i$ is the number of its neighbours with the
same strategy times the colour's prestige~$p(i) \geq 0$, i.e.~$u_i(s) \defas p(i) \abs{\set{j \in \Neighb(i) | s_i = s_j}}$.
Assume that $0 \in S_i, \forall i$.
\end{defin}
Intuitively, each player aims to be as coordinated as possible with
her neighbours, and on the most prestigious colour possible.

We consider the direct control set problem where the target equilibrium is $t = (0, 0, \ldots, 0)$.

\begin{theorem}\label{The:dec_cont_set_npc}
The decision versions of the direct control set
problems in a coordination game are \NP-complete.
\end{theorem}
\begin{proof}
We have proven the belonging to \NP{} even for the general problems,
so it remains to demonstrate the \NP-hardness.

We reduce the decision version of dominating set to the decision version of direct control set
on graphs. Recall that
\begin{defin}
Given an undirected graph $G = (V, E)$, a \defined{dominating set} is $D \subseteq E$, such that for any $x \in E \setminus D$, there exists $y \in D$ such that $(y, x) \in D$.

The decision version of \defined{Dominating Set} receives a graph $G = (V, E)$ and an integer~$k$ as input, and the decision is whether there exists a dominating set $D$ of size at most $k$.
\end{defin}
Given an instance of dominating set
$G = (V = \set{1, \ldots, n}, E)$, define the following coordination game.
First, let the bounds for the sizes of the dominating set of the direct control set decision are the same.

Make two copies of the original graph, where both copies of
each vertex are connected among themselves and to all the 
copies of the neighbours of that vertex in the original graph.
Let each copy be controlled by a separate player, thus having
a game with $2n$ players. Let the strategy set of
vertex~$i \in V$ be $S_i \defas \set{i, 0}$, and let the prestige be $p(i) \defas 1$, for every colour~$i$. The problem is to directly bring
the equilibrium~$s$ where every node~$i$ chooses $i$ to the
equilibrium~$d$ where every node chooses $0$.
In~$s$, deviating to $0$ is strictly worse that keeping playing $i$, 
because there is one neighbour playing $i$ and none are playing $0$.
Bringing player~$i$ from $s$ to $d$ means choosing the node or making at least one of its neighbours have the colour~$d_i$, which means a set is a
direct control set if and only if it is a dominating set in the original
graph.
\end{proof}

Since the decision problem is \NPH,
let us consider solving the search problem approximately.
When all the players choose between the same two options, namely
when $S_1 = \ldots = S_n = \set{0, 1}$, we reduce the 
direct control set problem to the
weighted $k$-dominating set problem, where we need at least $k$ neighbours
from the chosen set in the neighbourhood of every non-chosen vertex. 
That problem can be approximated~\cite{KuhnMoscibrodaWattenhofer2006},
so this reduction 

Our reduction works as follows:
\begin{enumerate}
	
	\item	For each vertex $i$, let $k_i$ be the minimum number
	of neighbours required to make $0$ become $i$'s best response.
	
	\item	Add $k - k_i$ unique zero-weighted neighbours to each original
	vertex. Make each such vertex play $0$.
	//Now, $i$ can require at least $k$ vertices, since it has $k - k_i$ 
	//for free.
\end{enumerate}

The case with more than two strategies cannot be dealt with by analogy,
because the minimum number
of neighbours player~$i$ needs to have $0$ as a best response depends
on which strategies each neighbours played before deviating.

%% file: conclusion.tex
\section{Conclusions and Future Work}\label{Sec:conclusion}

Having proven the safer problem from Definition~\ref{def:dir_control_set_prob:ord_ind}, where the control holds regardless of the order of the reaction, is hard even to verify, we turn to the direct control set Problem from Definition~\ref{def:dir_control_set_prob}. Proving it is excruciatingly hard even to approximate, we provide exact and approximate algorithms under certain assumptions, being most successful with potential games. 
The hardness result explain why we rarely observe practical influence by computing the best direct control set and pressurising those players to act out their desired strategies. Nonetheless, the success in some classes suggests where we can apply such methods, in addition to very small-scale interactions, where exhaustive search is reasonable.

We proved hardness of even verifying the solution to the order-independent direct control set problem. We should thus define and study other variations of the direct control set problem, aiming to strike the balance between being realistic and doable. For instance, for singleton congestion games, given any direct control set, all the other players keep being motivated to follow suit in the process of moving to the destination equilibrium in any order; however, such a path does not have to exist for a general congestion game.
We can also deal with the complexity barrier by heuristically improving the incremental Algorithm~\ref{alg:perm_add}, such as
picking the best orders of adding players to obtain a direct control set.

Furthermore, characterising the equilibria possessing direct control sets of a certain size is theoretically interesting.

Another direction relates to incentivising the direct control set players to play the strategies we want them to play. 
When everyone plays the original profile, being a control set member 
may naturally be non-profitable, and even after everybody has settled on the 
goal equilibrium, some players may have been better off in the original 
profile, even if the social welfare has increased. Being in 
the control set can thus be
subsidised, like in the mediated equilibrium~\cite{MondererTennenholtz09},
or connected to another game, played in parallel by those players.
When subsidised, the subsidy should be related to the actual loss
after the desired equilibrium is played and to the contribution of
each player to the control set, using cooperative game theory.
%

%
In summary, we provide a novel theoretical view to influencing play, while also providing practical algorithms for certain games.